\documentstyle[12pt,graphicx]{article}
\begin{document}
\title{Vacuum Polarization Effects in the Global Monopole Spacetime
in the Presence of Wu-Yang Magnetic Monopole}
\author{E. R. Bezerra de Mello \thanks{E-mail: emello@fisica.ufpb.br}
\\
Departamento de F\'{\i}sica-CCEN\\
Universidade Federal da Para\'{\i}ba\\
58.059-970, J. Pessoa, PB\\
C. Postal 5.008\\
Brazil}
\maketitle
\begin{abstract}
In this paper we consider the presence of the Wu-Yang magnetic monopole
in the global monopole spacetime and their influence on the vacuum
polarization effects around these two monopoles placed together. According
to Wu-Yang [Nucl. Phys. {\bf B107}, 365 (1976)] the solution of the 
Klein-Gordon equation in such an external field will not be an ordinary
function but, instead, {\it section}. Because of the peculiar radial 
symmetry of the global monopole spacetime, it is possible to cover its
space section by two overlapping regions, needed to define the singularity
free vector potential, and to study the quantum effects
due to a charged scalar field in this system. In order to develop this
analysis we construct the explicit Euclidean scalar Green {\it section} 
associated with a charged massless field in a global monopole spacetime 
in the presence of the Abelian Wu-Yang magnetic monopole. 
Having this Green section it is possible to study the vacuum polarization 
effects. We explicitly calculate the renormalized vacuum expectation value 
$\langle \Phi(x)^*\Phi(x) \rangle_{Ren.}$, associated with a
charged scalar field operator and the respective energy-momentum tensor, 
$\langle T_{\mu\nu}(x)\rangle_{Ren.}$, which are expressed in terms of the 
parameter which codify the presence of the global and magnetic monopoles.
\end{abstract}

\newpage
\renewcommand{\thesection}{\arabic{section}.}
\section{Introduction}
In this paper we calculate the Euclidean scalar Green section associated 
with a charged massless field in the global monopole spacetime in presence
of an Abelian Wu-Yang magnetic monopole. 

As it is well known the Wu-Yang magnetic monopole  \cite{W-Y} does not
possess strings of singularities around it. The vector potential $A_\mu$
is defined in two overlapping region $R_a$ and $R_b$ which cover the whole
space. Using spherical coordinates system $r$, $\theta$ and $\phi$ with the
monopole at origin we choose:
\begin{eqnarray}
\label{Ra}
R_a&:& 0 \leq \theta < \frac12\pi+\delta, \  r>0, \  0\leq \phi 
<2\pi \ , \\
R_b&:& \frac12\pi-\delta< \theta \leq \pi, \ r>0, \ 0\leq \phi < 2\pi ,\\ 
\label{Rab}
R_{ab}&:& \frac12\pi-\delta< \theta< \frac12\pi+\delta, \ r>0,  \
0\leq \phi < 2\pi \ , 
\end{eqnarray}
with $0<\delta \leq\frac12\pi$. $R_{ab}$ is the overlapping region.

The non vanishing components of vector potentials are
\begin{eqnarray}
\label{A}
(A_\phi)_a&=&g(1-\cos\theta) \ ,
\nonumber\\
(A_\phi)_b&=&-g(1+\cos\theta) \ ,
\end{eqnarray}
 being $g$ the monopole strength. In the overlapping region the non
vanishing components are related by a gauge transformation
\begin{equation}
(A_\phi)_a=(A_\phi)_b+\frac ieS_{ab}\partial_\phi S_{ab}^{-1} \ ,
\end{equation}
where $S=S_{ab}=e^{2iq\phi}$ being $q=eg=n/2$ in units $\hbar=c=1.$

The global monopole is a heavy object formed in the phase transition of a
system composed of a self-coupling scalar field triplet $\phi^a$ whose original
global $O(3)$ symmetry is spontaneously broken to $U(1)$ \cite{BV}.
The Euclidean version of the metric tensor associated with a pointlike global 
monopole can be given by the line element below:
\begin{equation}
\label{GM}
ds^2=d\tau^2+\frac{dr^2}{\alpha^2}+r^2d\Omega_{2}^2=g_{\mu\nu}(x)
dx^\mu dx^\nu \ .
\end{equation}
The parameter $\alpha$ which codify the presence of the global monopole is 
smaller than unity. This spacetime it is not flat: the scalar curvature 
$R=2(1-\alpha^2)/r^2$, and the solid angle associated with a sphere 
with unity radius is $\Omega=4\pi\alpha^2$, so smaller than ordinary one. 
The energy-momentum tensor associated with this object has a diagonal form 
and its non-vanishing components read $T_0^0=T_1^1=(\alpha^2-1)/r^2$.

Here in this paper we shall consider the quantum analysis of a massless
charged scalar field in the spacetime of a global monopole and in the
presence of a magnetic monopole. Specifically we are interested in 
calculation of the vacuum polarization effect on this field. 
In order to develop this analysis we shall consider that both monopoles, 
the global and the magnetic ones, are at 
the same position which we take as the origin of our reference system.
Doing this we try to reproduce the quantum analysis of a charged scalar
field in the presence of a more general object which presents the
main aspects of both monopoles: a solid angle metric spacetime
and a radial magnetic field. The complete treatment about this system
must take into account the contribution on the geometry of the spacetime
due to the energy density associated with the magnetic field \footnote{The
metric tensor associated with a pointlike compost topological object which
takes into account the presence of a magnetic charge in a solid angle
deficit geometry, has been obtained recently \cite{Mello}. The structure
of the respective manifold corresponds to a Reissner-Nordstr\"{o}m 
spacetime with a solid angle deficit factor. For large distance, the 
contribution due to the magnetic field can be neglected.}. However at this 
first moment, seems appropriate to us to start this analysis in a simpler 
way, i.e., considering only the influence of the global monopole on the 
geometry of the spacetime. 

As in the flat case, it is possible to cover the space section of the global
monopole spacetime into two overlapping regions $R_a$ and $R_b$ as in 
(\ref{Ra}) - (\ref{Rab}), with the metric tensor given by the same expressions 
as before in both regions, and define a vector potential 
as in (\ref{A}). In this way the magnetic field is expressed by the same 
expressions in both regions: $\vec{B}=g\hat{r}/r^2$.

The Klein-Gordon equation associated with a massless charged field in 
curved spacetime and in presence of an external four-vector potential $A_\mu$,
can be obtained from the covariant form of the Klein-Gordon equation replacing
the partial derivative $\partial_\mu$ by extended derivative $D_\mu=
\partial_\mu-ieA_\mu$:  
\begin{equation}
\label{KG}
\left[\Box-\frac{ieA^\mu}{\sqrt{g}}\left(\partial_\mu\sqrt{g}\right)-
ie\left(\partial_\mu A^\mu\right)-2ieA^\mu\partial_\mu-e^2A^\mu A_\mu+
\xi R\right]\Psi(x)=0 \ ,
\end{equation}
with
\begin{equation}
\Box\Psi=\frac1{\sqrt{g}}\partial_\mu\left[\sqrt{g}g^{\mu\nu}\partial_\nu
\Psi\right] .\
\end{equation}
In the above equation we have introduced a non-minimal coupling between the 
field with the geometry, $\xi R$, where $\xi$ is an arbitrary numerical
factor and $R$ is the Ricci scalar curvature. Being the four-vector 
representing the Wu-Yang magnetic monopole, this equation must be analysed in 
both regions, $R_a$ and $R_b$, separately. The solution $\Psi$ is not an 
ordinary function but, instead, a {\it section} assuming values $\Psi_a$ and 
$\Psi_b$ in $R_a$ and $R_b$, respectively, and satisfying the gauge 
transformation
\begin{equation}
\Psi_a=S_{ab}\Psi_b
\end{equation}
in the overlapping region $R_{ab}$.

In order to analyse this Klein-Gordon equation we shall adopt the usual 
approach: We write the solution in the form $\Psi(x)=
e^{-iEt}R(r)\Theta(\theta)\Phi(\phi)$, where the 
angular part  can be expressed in terms of the monopole harmonics, 
$Y_{lm}^q(\theta,\phi)$ \cite{W-Y1}. The monopole harmonics are eigensections
satisfying the eigenvalues equations below:
\begin{equation}
{\vec{L}_q}^2Y_{lm}^q=l(l+1)Y_{lm}^q \ ,
\end{equation}
and
\begin{equation}
L_zY_{lm}^q=mY_{lm}^q
\end{equation}
with $l=|q|, |q|+1,|q|+2, ...$ and $m=-l,-l+1, ... , l$. Being
\begin{equation}
\vec{L}_q=\vec{r}\times(\vec{p}-e\vec{A})-q\hat{r} \ .
\end{equation}

The temporal and radial functions, solutions of (\ref{KG}), have the same 
expressions in whole space.

In  section $2$ we derive the Euclidean Green section associated with
this system. In fact this Green section is given in terms of an infinity sum
of continuous section. In section $3$ we compute explicitly the 
renormalized vacuum expectation value $\langle\Phi^*(x)\Phi(x)\rangle$
of a charged massless scalar field operator for the case where 
$\eta^2=1-\alpha^2 \ll 1$ and also present a formal expression to 
$\langle T_\mu^\nu(x)\rangle_{Ren.}$. The scale dependent term in
$\langle T^\nu_\mu\rangle$ is explicitly calculated up to the first order in 
$\eta^2$ and $q^2$. In section $4$ we present our conclusions and the most
important remarks about this system.

\section{Green Section}

The Euclidean Green section associated with a charged massless scalar
field in the global monopole spacetime in presence of the Wu-Yang magnetic
monopole, can be obtained by the Schwinger-DeWitt formalism as follows:
\begin{equation}
\label{K}
G_E(x,x')=\int_0^\infty ds K(x,x';s) \ ,
\end{equation} 
where the heat kernel, $K(x,x';s)$, can be expressed in terms of the
eingensections of the Klein-Gordon operator defined in (\ref{KG}) as
shown below:
\begin{equation}
K(x,x';s)=\sum_\sigma \Psi_\sigma(x)\Psi_\sigma^*(x')e^{-s\sigma^2} \ ,
\end{equation}
$\sigma^2$ being the corresponding positively defined eigenvalue. Because of
the completeness of the monopole harmonics \cite{W-Y1}, the respective
normalized eigensection is:
\begin{equation}
\Psi_\sigma^q(x)=\sqrt{\frac{\alpha p}{2\pi}}e^{-i\omega\tau}\frac{J_{\lambda_l
}(pr)}{\sqrt{r}}Y_{lm}^q(\theta,\phi) \ ,
\end{equation}
with
\begin{equation}
\sigma^2=\omega^2+\alpha^2p^2 \ .
\end{equation}
$J_\lambda$ is the Bessel function of order
\begin{equation}
\label{L}
\lambda_l=\alpha^{-1}\sqrt{(l+1/2)^2-q^2+2(\alpha^2-1)(\xi+1/8)} \ .
\end{equation}
So according to (\ref{K}) the heat kernel is given by
\begin{eqnarray}
\label{K1}
K(x,x';s)&=&\int^\infty_{-\infty}d\omega\int^\infty_0 dp\sum_{l,m}
\Psi_\sigma(x)\Psi^*_\sigma(x')e^{-s\sigma^2}
\nonumber\\
&=&\frac1{4\alpha s^{3/2}}\frac1{\sqrt{\pi rr'}}
e^{-\frac{\Delta\tau^2\alpha^2+r^2+r'^2}{4\alpha^2s}}\times
\nonumber\\
&&\sum_{l=|q|}^\infty I_{\lambda_l}\left(\frac{rr'}{2\alpha^2 s} 
\right)\sum_{m=-l}^lY_{lm}^q(\theta,\phi)
\left(Y_{lm}^q(\theta',\phi')\right)^* \ ,
\end{eqnarray}
$I_\lambda$ being the modified Bessel function. Tai Tsun Wu and Chen Ning
Yang in \cite{W-Y2} have derived some properties of monopole harmonics, 
including the generalization of spherical harmonics addition theorem. However, 
because we are interested to calculate the
renormalized value of the Green function in the coincidence limit, a
simpler expression is obtained taking $\theta=\theta'$ and $\phi=\phi'$
in (\ref{K1}) \footnote{In Appendix A we show the simplified result to the
sum of product of monopole harmonics in (\ref{Green}) when we take the 
coincidence limit in the angular variables.}. Moreover, substituting this 
simplified expression into (\ref{K}) we obtain, with the help of \cite{G}, the 
following Green function:
\begin{equation}
\label{Green}
G^q(r,\tau;r',\tau')=\frac1{8\pi^2rr'}\sum_{l=|q|}^\infty(2l+1)
Q_{\lambda_l-1/2}\left(\frac{\alpha^2\Delta\tau^2+r^2+r'^2}{2rr'}\right) \ ,
\end{equation}
$Q_\lambda$ being the Legendre function. Unfortunately because of the 
non-trivial dependence of the order of the Legendre function with the parameter 
$\alpha$ and $q$, it is not possible to obtain a closed 
expression to the above Green section, even for the case $\xi=-1/8$. The 
best that we can do is to develop an approximated expression to it.
This development will be presented in the next subsection. 

Having now this Green section it is possible to obtain the vacuum polarization 
effect in this gravitational background in the presence of the Wu-Yang 
magnetic monopole. 

\subsection{Computation of $\langle\Phi^*(x)\Phi(x')\rangle_{Ren.}$}

Here in this subsection we shall develop the calculation of the vacuum 
expectation value of the square of the modulus of the field operator. This 
quantity can be obtained calculating the Green section in the coincidence
limit as shown below:
\begin{equation}
\langle\Phi^*(x)\Phi(x)\rangle=\lim_{x' \to x}G^q(x,x') \ .
\end{equation}
However this procedure provides a divergent result. In order to obtain
a finite and well defined result we must apply in this calculation some
renormalization procedure. Here in this paper we shall adopt the
point-splitting renormalization one. This procedure is based upon a divergence
subtraction scheme in the coincidence limit of the Green function. 
Adler {\it et al} \cite{Adler} have pointed out that the singular behavior of 
the Green function has the same structure as the Hadamard function. So in order
to obtain a finite result to the vacuum expectation value above, we
subtract from the Green section the Hadamard function, before applying the
coincidence limit: \footnote{In the calculation of renormalized vacuum
expectation value of the energy-momentum operator, $\langle T_{\mu\nu}
\rangle$, Wald \cite{Wald} has detected an error in \cite{Adler} and modified 
the prescription adding an extra term proportional to $a_2(x)$ in order
to provide the correct result for the trace anomaly.} 
\begin{equation}
\label{Psi}
\langle\Phi^*(x)\Phi(x)\rangle_{Ren.}=\lim_{x' \to x}\left[G^q(x,x')-
G_H(x,x')\right] \ .
\end{equation} 
  
In order to develop this calculation, let us take first in (\ref{Green})
$\Delta\tau=0$ and use for the Legendre function its integral representation 
\cite{G}: 
\begin{equation}
Q_{\lambda-1/2}(\cosh \rho)=\frac1{\sqrt{2}}\int_\rho^\infty dt
\frac{e^{-\lambda t}}{\sqrt{\cosh t- \cosh \rho}} \ ,
\end{equation}
with $\cosh \rho=\frac{r^2+r'^2}{2rr'}$. So, substituting this expression into
(\ref{Green}) we get:
\begin{equation}
\label{Gq}
G^q(r,r')=\frac1{8\pi^2rr'}\frac1{\sqrt{2}}\int_\rho^\infty dt
\frac 1{\sqrt{\cosh t-\cosh \rho}}\sum_{l=|q|}^\infty (2l+1)e^{-\lambda_lt} \ .
\end{equation}
Unfortunately because of the dependence of $\lambda_l$ with $\alpha^2$ and 
$q^2$, it is not possible to develop the summation above and to 
express (\ref{Gq}) in a closed form. So, in order to obtain a more workable 
expression, it is necessary to make some approximations: The first one is to 
consider the parameter $\eta^2=1-\alpha^2 \ll 1$ \footnote{In fact for a 
typical grand unified theory where this topological defect appears as a 
consequence of spontaneously break of the global symmetry, $\eta^2$ is of 
order $10^{-6}$} and to develop a series expansion in powers of this parameter.
The second procedure to be adopted, is to develop again another expansion in 
the resulting expressions in powers of $\frac{q^2}{(l+1/2)^2}$. In this way 
for each order in $\eta^2$ in the summation, we procedure another expansion. 
After doing this it is possible to develop the summation. Our approximated 
expression was developed only up to the first order in 
$\eta^2$ and the next expansion, which involves $\sqrt{1-q^2/(l+1/2)^2}$, 
was developed up to the first order too.  The final result presents also an 
exponential dependence on $q$, $e^{qt}$, due the fact that all terms in 
the summations start from $l=q$ \footnote{From now on in this paper we 
shall consider $q\geq 0$. The extension of our results to negative values of
$q$ are straightforward.}.

The summation above can be written as having two mains contributions as 
shown below:
\begin{equation}
S=\sum_{l=q}^\infty (2l+1)e^{-\lambda_l t}\simeq S_1-\frac{\eta^2 t}2S_2 \,
\end{equation}
with
\begin{equation}
S_1=\frac{e^{-qt}}{2\sinh(t/2)}\left[\coth(t/2)+2q\right]+\frac{q^2e^{-qt}t}
{2\sinh(t/2)} \ ,
\end{equation}
and after some calculations,
\begin{eqnarray}
S_2&=&\frac{e^{-qt}}{8\sinh^2(t/2)}\left[8q\cosh(t/2)+8q^2\sinh(t/2)+
\frac{3+\cosh(t)}{\sinh(t/2)}\right]
\nonumber\\
&+&\frac{q^2e^{-qt}}{2\sinh(t/2)}\left[\frac t2\coth(t/2)-1\right]-
2(\xi+1/8)e^{-qt}\frac1{\sinh(t/2)}\times
\nonumber\\
&&\left[1+q^2te^{-t/2}\sinh(t/2)\Phi(e^{-t},1,q+1/2)+\frac{4q^2}{(2q+1)^2}
e^{-t/2}\right.\times
\nonumber\\
&&\left.\sinh(t/2)_3F_2(1,1/2+q,1/2+q;3/2+q,3/2+q;e^{-t})\right] \ ,
\end{eqnarray} 
where $_3F_2$ is an hypergeometric function and $\Phi(z,s,v)$ the Phi-function
\cite{G}.
  
The Hadamard function in a curved four dimensional spacetime contains a
logarithmic term which depends on the arbitrary cutoff scale $\mu$ and is
expressed in terms of the square of the geodesic distance $2\sigma(x,x')$ 
as shown below:
\begin{equation}
G_H(x,x')=\frac1{16\pi^2}\left[\frac2{\sigma(x,x')}-(\xi+1/6)R
\ln\left[\frac{\mu^2}2\sigma(x,x')\right]\right] \ .
\end{equation}
For the radial point splitting we have $\sigma (x,x')=\frac{(r-r')^2}
{2\alpha^2}$.

The above expression can be written in a more convenient way expressing
$1/(r-r')^2$ in terms of an integral similar to that one which appears
in (\ref{Gq}), and the logarithm in terms of $Q_0(\cosh\rho)$. So, the final
expression is
\begin{eqnarray}
G_H(r,r')&=&\frac{1-\eta^2}{16\pi^2}\frac1{rr'\sqrt{2}}\int_\rho^\infty dt
\frac1{\sqrt{\cosh t-\cosh \rho}}\frac{\cosh(t/2)}{\sinh^2(t/2)}
\nonumber\\
&+&\frac1{4\pi^2\sqrt{2}}(\xi+1/6)\frac{\eta^2}{r^2}\int_\rho^\infty dt
\frac1{\sqrt{\cosh t-\cosh \rho}}e^{-t/2}
\nonumber\\
&-&\frac1{8\pi^2}(\xi+1/6)\frac{\eta^2}{r^2}\ln\left[\frac{\mu^2(r+r')^2}
{4\alpha^2}\right] \ .
\end{eqnarray}

Finally below we present the renormalized vacuum expectation value
associated with the square of the modulus of the scalar field. Because 
this expression is a long one, we shall present it written in terms of four 
different contributions: The purely magnetic part, followed by two other 
proportional to the parameter $\eta^2$ which also depend on $q$. The last 
contribution is the cutoff dependent term which disappears when we take 
$\xi=-1/6$: 
\begin{eqnarray}
\label{Gq1}
\langle\Phi^*(x)\Phi(x)\rangle_{Ren.}&=&\frac1{32\pi^2r^2}I_1(q)-\frac{\eta^2}
{32\pi^2r^2}I_2(q)+\frac{\eta^2}{32\pi^2r^2}(\xi+1/8)I_3(q)
\nonumber\\
&+&\frac{\eta^2}{4\pi^2r^2}(\xi+1/6)\ln(\mu r) \ ,
\end{eqnarray}
where
\begin{eqnarray}
I_1(q)&=&\int_0^\infty dt\frac1{\sinh(t/2)}\left[\frac{e^{-qt}}{\sinh(t/2)}
[\coth(t/2)+2q]+\frac{q^2e^{-qt}}{\sinh(t/2)}\right.
\nonumber\\
&-&\left.\frac{\coth(t/2)}{\sinh(t/2)}
\right] \ ,
\end{eqnarray}
\begin{eqnarray}
I_2(q)&=&\int_0^\infty dt\frac1{\sinh(t/2)}\left[\frac{te^{-qt}}{8\sinh^2(t/2)}
\left[8q\cosh(t/2)+8q^2\sinh(t/2)\right.\right.
\nonumber\\
&+&\left.\frac{3+\cosh(t)}{\sinh(t/2)}\right]+\frac{q^2te^{-qt}}{2\sinh(t/2)}
[t/2\coth(t/2)-1]-\frac{\coth(t/2)}{\sinh(t/2)}
\nonumber\\
&+&\left.\frac{e^{-t/2}}6\right] \ ,
\end{eqnarray}
and
\begin{eqnarray}
I_3(q)&=&\int_0^\infty dt\frac1{\sinh(t/2)}\left[\frac{2te^{-qt}}{\sinh(t/2)}
\left[1+q^2te^{-t/2}\sinh(t/2)\times\right.\right.
\nonumber\\
&&\Phi(e^{-t},1,q+1/2)+\frac{4q^2}{(2q+1)^2}e^{-t/2}\sinh(t/2)\times
\nonumber\\
&&\left.\left._3F_2(1,1/2+q,1/2+q;3/2+q,3/2+q;e^{-t})\right]-
4e^{-t/2}\right] \ .
\end{eqnarray}

From the above expressions it is possible to observe that all the 
integrand are regular at $t \to 0$ and vanish for $t \to
\infty$. This is a consequence of all the divergences of $G^q(x,x')$ 
in the coincidence limit, do not depend on the parameter $q$. 
On the other hand in $(1+3)-$dimensional spacetime the Hadamard function 
does not depend on the gauge field, consequently if we had developed an
expansion in higher powers of $\frac{q^2}{(l+1/2)^2}$ in the
summation of (\ref{Gq}), all these corrections, besides to be subdominant,
would provide finite results to the renormalized vacuum expectation 
value (\ref{Psi}).\footnote{For higher even dimensional spacetime as $(1+5)$ 
case, the Hadamard function depends on the coefficient $a_2(x)$, which by its 
turn depend on the gauge field Lagrangian density \cite{Ian}.} 

The calculation of the renormalized vacuum expectation value associated with
a massless scalar field in the global monopole spacetime has been calculated 
a few years ago by Mazzitelli and Lousto \cite{ML}. From our results 
it is possible to see that taking $q=0$ into (\ref{Gq1}), the integral
$I_1(q)$ vanishes and we obtain the same expression as found in equation 
$(2.17)$ of \cite{ML}. 

The complete form for $I_1(q)$ and $I_2(q)$ can be obtained analytically
in terms of $q$ using the computer program MAPLE; however they are very 
long expressions. Unfortunately as to $I_3(q)$, we could not find the 
respective integral. Because this result is one of the most important one
of this paper we decided to evaluate all the integral for specific values of 
the parameter $q$: $q=1/2$ and $q=1$.  For theses two values the two first
integrals acquire a closed and short results, however $I_3$ can only be
evaluated numerically. In the Appendix B we present more workable expressions
to $\Phi(e^{-t},1,q+1/2)$ for the case of interest when $q$ is an integer
or half-integer. Our numerical results are given below:
\noindent\\
For $I_1(q)$: 
\begin{equation}
q=1/2, \ I_1=-\frac{\pi^2}4+\frac32 \ ,
\end{equation}
\begin{equation} 
q=1, \ I_1=-\frac{2\pi^2}3+4 \ . 
\end{equation}
For  $I_2(q)$:
\begin{equation}
q=1/2, \ I_2=\frac1{144}(44+126\zeta(3)-96\ln(2)-18\pi^2) \ ,
\end{equation}
\begin{equation}
q=1, \ I_2=\frac1{18}(16+36\zeta(3)-9\pi^2) \ .
\end{equation}
For $I_3(q)$:
\begin{equation}
q=1/2, \ I_3 \approx 1.46785 \ 
\end{equation}
and
\begin{equation}
q=1, \ I_3 \approx -2.12959 \ .
\end{equation}

\subsection{Dimensional Arguments of $\langle T_{\mu\nu} \rangle$}

In this paper we are analysing under a quantum point of view, the behavior
of a massless charged scalar field in the pointlike global monopole spacetime, 
defined by the metric tensor given in (\ref{GM}), in the presence of a 
magnetic field defined by (\ref{A}). In previous subsection we have
obtained the renormalized vacuum expectation value of the square of the modulus
of the scalar quantum field. Our result shows an explicit dependence of this
term with the two fundamental dimensionless parameters, $\alpha$ and $q$, and 
also with radial coordinate $r$ and the renormalization mass scale $\mu$. 

Although we are not going to develop the explicit calculation of the 
renormalized vacuum expectation value of the operator energy-momentum tensor 
associated with this system, $\langle T_{\mu\nu}(x) \rangle_{Ren.}$, we 
expect by dimensional arguments only that this quantity be proportional to $
1/r^4$ \footnote{This dependence is a consequence of the fact that there is 
no other dimensional parameter, since we are working with natural units 
$\hbar=c=1$.}. The factor of proportionality should be given in terms of 
$\alpha$, $q$, $\xi$ and also depends on the renormalization mass
scale. So briefly speaking we can say that 
\begin{equation}
\langle T^\nu_\mu(x)\rangle_{Ren.}=\frac{G^\nu_\mu(q,\alpha,\xi,\mu r)}
{r^4} \ .
\end{equation}
On the other hand, by symmetry of this model, the above vacuum expectation
value should be diagonal. Moreover this quantity must be conserved,
\begin{equation}
\nabla_\nu \langle T^\nu_\mu(x)\rangle_{Ren.}=0 \ ,
\end{equation}
and satisfies the trace anomaly
\begin{equation}
\langle T^\mu _\mu(x)\rangle_{Ren.}=\frac1{16\pi^2}a_2(x) \ .
\end{equation}

In order to explore the above conditions in a more appropriate way, let us
write the vacuum expectation value by 
\begin{equation}
\langle T^\nu_\mu(x)\rangle_{Ren.}=\frac1{16\pi^2r^4}\left[A^\nu_\mu(q,\alpha,
\xi)+B^\nu_\mu(q,\alpha,\xi)\ln(\mu r)\right] \ ,
\end{equation}
where $A^\nu_\mu$ and $B^\nu_\mu$ in principal are arbitrary numbers. However
because the cutoff factor $\mu$ is completely arbitrary, there is an ambiguity 
in the definition of this renormalized vacuum expectation value. Moreover the 
change in this quantity under a change of the renormalized scale is given in 
terms of the tensor $B^\nu_\mu$ as shown below:
\begin{equation}
\langle T^\nu_\mu(x)\rangle_{Ren.}(\mu)-\langle T^\nu_\mu(x)\rangle_{Ren.}
(\mu')=\frac1{16\pi^2r^4}B^\nu_\mu(q,\alpha,\xi)\ln(\mu/\mu') \ .
\end{equation}
On the other hand, Christensen in his beautiful paper \cite{Crhistensen}
has presented a general expression for this difference in terms of the 
variation of the effective action which depends on the logarithmic term: 
it can be calculated by
\begin{equation}
\langle T^\nu_\mu(x)\rangle_{Ren.}(\mu)-\langle T^\nu_\mu(x)\rangle_{Ren.}
(\mu')=\frac1{16\pi^2}\frac1{\sqrt{g}}\frac{\delta}{\delta g^{\mu\nu}}
\int d^4x\sqrt{g} a_2(x)\ln(\mu/\mu') \ ,
\end{equation}
where $a_2(x)$, which is fourth order term in derivative of the metric
tensor, also presents contribution from the gauge field strength as show
below \cite{Ian}:
\begin{eqnarray}
a_2(x)&=&-\frac1{180}R_{\alpha\beta\gamma\delta}R^{\alpha\beta\gamma\delta}+
\frac1{180}R_{\alpha\beta}R^{\alpha\beta}+\frac16\left(\frac15-\xi\right)
\Box R
\nonumber\\
&-&\frac12\left(\frac16-\xi\right)^2R^2-\frac1{12}\omega_{\alpha\beta}
\omega^{\alpha\beta} \ .
\end{eqnarray}

For the global monopole spacetime in the presence of the Wu-Yang magnetic
monopole, the above coefficient can be explicitly obtained by using the 
only non vanishing components of the vector potential (\ref{A}) and the 
Riemann tensors $R_{2323}=R_{3232}=(1-\alpha^2)r^2\sin^2\theta$. 
After some intermediate steps we get:
\begin{equation}
a_2(x)=\frac{1-\alpha^2}{r^4}\left[(\alpha^2-1)\left(\frac1{90}+
2\left(\frac16-\xi \right)^2\right)+\frac23\left(\frac15-\xi \right)\alpha^2
\right]+\frac{q^2}{6r^4} \ .
\end{equation}

The geometric part of the tensor $B^\nu_\mu$ associated with the pointlike 
global monopole, has been obtained in \cite{ML}. So the new contribution here
is due to the field strength associated with the Wu-Yang magnetic monopole.
After some intermediate steps we arrive, up to the first order
in the $\eta^2$ parameter, to
\begin{eqnarray}
B^\nu_\mu(q,\eta^2,\xi)&=&\frac1{90}\eta^2 diag(1,1,-1,-1)+8\eta^2(\xi+1/6)^2
diag(-1/2,1,-1,-1)
\nonumber\\
&-&\frac{q^2}{12}diag(1,1,-1,-1) \ ,
\end{eqnarray}
which has the same expression in whole space.

So for $\xi=-1/6$ the trace anomaly, up to the first order in $\eta^2$, reads:
\begin{equation}
\langle T^\mu_\mu(x)\rangle_{Ren.}=\frac{\eta^2}{720\pi^2 r^4}+
\frac{q^2}{96\pi^2 r^4} \ .
\end{equation}

The conservation conditions on the renormalized vacuum expectation value
of the energy-momentum tensor imposes profound restrictions on the tensor 
$A^\nu_\mu$. After some calculations we arrive at:
\begin{equation}
A^0_0=A^1_1-B^1_1+T+(B^1_1-B^0_0)\ln(\mu r) \ ,
\end{equation}
and
\begin{equation}
A_3^3=A^2_2=-A_1^1+\frac12B^1_1-(B^1_1+B^2_2)\ln(\mu r) \ .
\end{equation}
When we take $\xi=-1/6$ we have
\begin{equation}
A_0^0=A_1^1+T-B^1_1 \ ,
\end{equation}
\begin{equation}
A_3^3=A_2^2=-A^1_1+\frac12B^1_1 \ ,
\end{equation}
where
\begin{equation}
T=r^4a_2(x)=\frac{\eta^2}{45}+\frac{q^2}6+O(\eta^4) \ .
\end{equation}
So the complete evaluation of $\langle T^\nu_\mu(x)\rangle_{Ren.}$ requires
the knowledge of at least one component of the tensor $A^\nu_\mu$, say
$A^1_1$. However this is a very long calculation which we do not attempt to
do here.

\section{Concluding Remarks}

In this paper we have analysed the vacuum polarization effect due to a 
massless charged scalar field in the idealized global monopole spacetime 
which presents a Wu-Yang magnetic monopole in its core. In order to develop 
this analysis we had to obtain the respective Green section, which corresponds 
to a generalization of the ordinary Green function having as its angular part 
the monopole harmonics.
This formalism was possible to be developed for this case because of the 
peculiar radial symmetry of the global monopole spacetime, once the Wu-Yang
magnetic monopole has been placed at its origin. In this way the two
overlapping regions $R_a$ and $R_b$ as in (\ref{Ra})-(\ref{Rab}), needed
to define the singularity free vector potential, can be established.

Considering the global monopole at the same position as the magnetic one, we 
tried to reproduce the vacuum polarization effect on a massless charged scalar
field due to a more general object which contains peculiar aspects of both
monopoles: a deficit solid angle spacetime and a radial magnetic field. We
are aware that if we had considered the energy density associated
with the magnetic field, the geometry of the spacetime would have some
contribution due to the latter. So the complete treatment for this problem 
must take into account both effects on the geometry.

In the calculation of $\langle \Phi^*(x)\Phi(x)\rangle_{Ren.}$ we had to
adopt two different approximations procedure: the first one was to develop
an expansion in the parameter $\eta^2=1-\alpha^2 \ll 1$ in the Green
{\it section}. Doing this expansion it was possible to obtain a more workable
expression. However we found another difficulty. The
remaining expressions did not allow us to proceed the summations on the
angular quantum number $l$. In this way we had to develop the second expansion
in $q^2/(l+1/2)^2$, where then all the summations were possible to be performed.
We decided to go in our approximations up to the first order only 
in both cases. Of course more precise expansions could be developed for both 
parameters. However these extra terms would be sub-dominant contributions in 
this specific calculation. 

Finally we want to say that in this paper we have considered the magnetic
field as an external one, i.e., we did not consider the influence of this
gauge field on the geometry of the spacetime. The exact solution of the
metric tensor associated with a pointlike non-Abelian magnetic monopole
has a Reissner-Nordstr\"{o}m form \cite{MM}. 

\newpage
{\bf{Acknowledgments}}
\\       \\
We would like to thank Conselho Nacional de Desenvolvimento Cient\'\i fico e 
Tecnol\'ogico (CNPq.) for partial financial support.

\section{Appendix A: The Addition Theorem}

Here in this appendix we present how the simplified expression to (\ref{Green})
is obtained when we take the coincidence limit in the angular variables in
(\ref{K1}).  In \cite{W-Y2} it is given the generalized monopole harmonics
addition theorem which we reproduce below with convenient modifications. For
both monopole harmonics defined in $R_a$ we have:
\begin{equation}
\sum_m Y^q_{l,m}(\theta, \phi)(Y^q_{l,m}(\theta',\phi'))^*=
\sqrt{(2l+1)/4\pi}Y^q_{l,-q}(\gamma,0)
e^{iq(\phi-\phi')}e^{-iq(R+R'-\pi)} \ ,
\end{equation}
where 
\begin{equation}
\cos\gamma=\cos\theta\cos\theta'+\sin\theta\sin\theta'\cos(\phi-\phi')
\end{equation}
and $R$ and $R'$ are angles defined in the Fig. 2 of \cite{W-Y2}. Taking
$\phi=\phi'$, we obtain $R'=0$ and $R=\pi$. On the other hand
\begin{equation}
Y^q_{l,-q}(\gamma,0)=\frac1{2^q}\sqrt{\frac{2l+1}{4\pi}}(1+x)^{-q}
\frac1{2^n n!}\frac{d^n}{dx^n}[(x-1)^n(x+1)^{n+2q}] \ ,
\end{equation}
with $x=\cos\gamma$ and $n=l-q$. Now taking $\gamma=0$ into the above
equation we get:
\begin{equation}
Y^q_{l,-q}(0,0)=\sqrt{\frac{2l+1}{4\pi}} \ .
\end{equation}

A similar result would be obtained if we have considered both monopole
harmonics in $R_b$. (Of course the situation where both monopole
harmonics are in different regions are not pertinent here.)

Another point which deserve to be mentioned in the generalized monopole
harmonics addition theorem is that the result is still a section. So
the name given to the Green {\it section} seems appropriate to us.

\section{Appendix B: Some Useful Identity}

In the numerical evaluation of $I_3(q)$ in (\ref{Gq1}) for specific values
of $q$ equal to $1/2$ and $1$, we needed an explicit expression for the 
$\Phi(e^{-t},1,q+1/2)$:
\begin{equation}
\Phi(e^{-t},1,q+1/2)=\sum_{n=0}^\infty \frac{e^{-nt}}{n+q+1/2} \ .
\end{equation}

Because $q$ assumes only integer or half-integer values this special function 
can be expressed by elementary functions as:
\noindent\\
For integer $q$,
\begin{equation}
\Phi=e^{qt}e^{t/2}\ln(\coth(t/4))-e^{qt}\sum_{n=0}^{q-1}\frac{e^{-nt}}
{n+1/2} \ ,
\end{equation}
and for half-integer $q$,
\begin{equation}
\Phi=-e^{[q]t}e^{t}\ln(1-e^{-t})-e^{[q]t}\sum_{n=0}^{[q]-1} \frac{e^{-nt}}
{n+1} \ ,
\end{equation}
where $[q]$ is the integer part of $q$.

\end{document}